\documentclass[conference, anonymous]{IEEEtran}
\IEEEoverridecommandlockouts
\usepackage{cite}
\usepackage{amsmath,amssymb,amsfonts}
\usepackage{algorithmic}
\usepackage{graphicx}
\usepackage{xspace}
\usepackage{textcomp}
\usepackage[table]{xcolor}

\usepackage{xcolor}

\def\BibTeX{{\rm B\kern-.05em{\sc i\kern-.025em b}\kern-.08em
    T\kern-.1667em\lower.7ex\hbox{E}\kern-.125emX}}
\usepackage[american]{babel}
\usepackage[utf8]{inputenc} 
\usepackage[scaled=0.92]{helvet}
\usepackage[nolist]{acronym}
\usepackage{amsfonts} 	% extra math fonts
\usepackage{amsmath} 	% mathematical formulas

\usepackage{amssymb} 	% Symbole für Zahlenmengen usw.
\usepackage{graphicx}
\usepackage{xcolor}
\usepackage{bookmark}
\usepackage{subcaption}
\usepackage{xspace}
\usepackage{tikz}
\usepackage{array}
\usepackage{multirow, makecell}
\usepackage{tabulary}
\usepackage{listings}
\usepackage{soul}
\usepackage{url}%
\usepackage{hyperref}
\usepackage{cleveref}
\nocite{*}

\newcommand\myshade{85}
\colorlet{mylinkcolor}{violet}
\colorlet{mycitecolor}{orange}
\colorlet{myurlcolor}{gray}

\hypersetup{
	linkcolor  = mylinkcolor!\myshade!black,
	citecolor  = mycitecolor!\myshade!black,
	urlcolor   = myurlcolor!\myshade!black,
	colorlinks = true,
}

\usepackage{pifont}% http://ctan.org/pkg/pifont
\definecolor{darkgreen}{rgb}{0.0, 0.5, 0.0}
\definecolor{bleudefrance}{rgb}{0.19, 0.55, 0.91}
\definecolor{bazaar}{rgb}{0.6, 0.47, 0.48}
\definecolor{atomictangerine}{rgb}{1.0, 0.6, 0.4}

\usepackage{listings}
\usepackage{lstautogobble}

\newcommand{\secucheck}{\textsc{SecuCheck}\xspace}
\newcommand{\secucheckkotlin}{\textsc{SecuCheck-Kotlin}\xspace}
\newcommand{\fluenttql}{\textsc{\textit{fluent}TQL}\xspace}
\newcommand{\fluenttqltoEng}{\textsc{\textit{fluent}TQL2English}\xspace}

\newcommand{\soot}{\textsc{Soot}\xspace}

\newcommand{\RQ}[1]{\textbf{RQ#1}\xspace} % research questions

\newacro{template}[UPB-CS-TT]{Paderborn University Computer Science thesis template}

\newcommand{\Cmicrobench}{Micro benchmark\xspace}
\newcommand{\microbench}{micro benchmark\xspace}

\newcommand{\kotlinsecucheck}{\textsc{SecuCheck-Kotlin}\xspace}

\newcommand{\jimplewrapper}{\textsc{JimpleProvider}\xspace}

\newcommand{\keyword}[1]{\texttt{#1}}

\newcommand{\funRef}[1]{\textit{\texttt{\sloppy{\fontfamily{qcr}\selectfont#1}}}}

\makeatletter
\newcommand{\customlabel}[2]{%
	\protected@write \@auxout {}{\string \newlabel {#1}{{#2}{\thepage}{#2}{#1}{}} }%
	\hypertarget{#1}{#2}
}
\makeatother

\newcommand{\nullable}[1]{\textcolor{BurntOrange}{\bfseries{?}}}

\definecolor{azure(colorwheel)}{rgb}{0.0, 0.5, 1.0}

\definecolor{codeIndent}{HTML}{CCCCCC}

\definecolor{aliceblue}{rgb}{0.94, 0.97, 1.0}

\newcommand{\constant}[1]{\sloppy{\fontfamily{qcr}\selectfont#1}}
\usepackage{adjustbox}
\definecolor{javared}{rgb}{0.6,0,0} % for strings
\definecolor{javagreen}{rgb}{0.25,0.5,0.35} % comments
\definecolor{javapurple}{rgb}{0.5,0,0.35} % keywords
\definecolor{javadocblue}{rgb}{0.25,0.35,0.75} % javadoc
\definecolor{javabg}{rgb}{0.95,0.95,0.92}

\lstdefinestyle{Java}
{
	language=Java,
	basicstyle=\fontsize{07}{08}\ttfamily,
	keywordstyle=\color{javapurple}\bfseries,
	stringstyle=\color{javared},
	commentstyle=\color{javagreen},
	morecomment=[s][\color{javadocblue}]{/**}{*/},
	numbers=left,
	numberstyle=\tiny\color{black},
	stepnumber=1,
	numbersep=5pt,
	tabsize=1,
	showspaces=false,
	showstringspaces=false,
	xleftmargin=10pt,
	autogobble=true,
	backgroundcolor=\color{javabg},
	framexleftmargin=1.5pt,
	framextopmargin=0.01mm,
	framexbottommargin=0.01mm, 
	frame=tb, framerule=0pt
}
\definecolor{annotationColor}{rgb}{0.72, 0.45, 0.2}
\definecolor{localVariables}{rgb}{0.8, 0.1, 1.0}

\begin{document}

\title{To what extent can we analyze Kotlin programs using existing Java taint analysis tools?\\(Extended Version)}

\author{\IEEEauthorblockN{Ranjith Krishnamurthy}
	\IEEEauthorblockA{\textit{Fraunhofer IEM} \\
		ranjith.krishnamurthy@iem.fraunhofer.de}
	\and
	\IEEEauthorblockN{Goran Piskachev}
	\IEEEauthorblockA{\textit{Fraunhofer IEM} \\
		goran.piskachev@iem.fraunhofer.de}
	\and
	\IEEEauthorblockN{Eric Bodden}
	\IEEEauthorblockA{\textit{Paderborn University \& Fraunhofer IEM} \\
		eric.bodden@uni-paderborn.de}
}

\maketitle

\begin{abstract}
As an alternative to Java, Kotlin has gained rapid popularity since its introduction and has become the default choice for developing Android apps. However, due to its interoperability with Java, Kotlin programs may contain almost the same security vulnerabilities as their Java counterparts. Hence, we question: \textit{to what extent can one use an existing Java static taint analysis on Kotlin code?} In this paper, we investigate the challenges in implementing a taint analysis for Kotlin compared to Java. To answer this question, we performed an exploratory study where each Kotlin construct was examined and compared to its Java equivalent. We identified 18 engineering challenges that static-analysis writers need to handle differently due to Kotlin's unique constructs or the differences in the generated bytecode between the Kotlin and Java compilers. For eight of them, we provide a conceptual solution, while six of those we implemented as part of \secucheckkotlin, an extension to the existing Java taint analysis \secucheck. 
\end{abstract}


\section{Introduction}
\label{sec:introduction}

Ten years since its introduction, Kotlin has been one of the fastest-growing programming languages (PLs). As of June 2022, it is the twelfth most popular PL by the PYPL index\footnote{\url{https://pypl.github.io/PYPL.html}}. Additionally, over 60\% of the Android apps are written in Kotlin, earning it the title of the default PL for the Android framework\footnote{\url{http://surl.li/cfrcc}}. One of the Kotlin advantages as a JVM-based PL is its interoperability with Java and its unique constructs like data classes, coroutines, null safety, extensions, etc. %Additionally, Kotlin introduces constructs that allow developers to write more compact and readable code. 

Like Java, Kotlin code may be vulnerable to security vulnerabilities, such as SQL injection~\cite{cwe89}. Therefore, statically analyzing Kotlin code can be a helpful method for detecting bugs and security vulnerabilities as early as possible. Despite its popularity, very few static-analysis tools can analyze Kotlin code, such as KtLint~\cite{ktlint}, Detekt~\cite{detekt}, Diktat~\cite{diktat}, and SonarQube~\cite{sonar}. These tools only perform pattern-based analyses using simple rules, such as the rules of SonarQube~\cite{sonarRules}. We are not aware of any tool that performs deep data-flow analyses on Kotlin code. For example, taint analysis has proven to be very useful for detecting many prevalent security vulnerabilities~\cite{fluenttql} such as injections~\cite{cwe89, cwe77, cwe476} and XSS~\cite{xss}. This versatility of the taint analysis is due to its capacity to set various inputs in the form of rules. At its core, the analysis follows the path between so-called sources, where the taint is created, until so-called sinks, where the taint is reported. The information for the sources and sinks is often encoded in a rule via a domain-specific language (DSL).

For Java, there are many existing taint analyses~\cite{flowdroid, secucheck} that can be used to detect many taint-style security vulnerabilities. Since Kotlin compiles to the Java bytecode, theoretically, one can use existing Java taint analyses on Kotlin code. However, the Kotlin compiler generates the bytecode differently than that of Java. This leads to the question: \textit{can one use taint analysis tools intended for Java to analyze Kotlin programs, or must one reinvent the wheel?}

In this paper, we report the result of an exploratory study that we conducted to address this question. We analyzed the Kotlin-generated bytecode for each language construct and compared it to the Java equivalent. We used the Jimple intermediate representation generated by the Soot framework~\cite{soot} for this comparison. For completeness, we used the official Kotlin documentation~\cite{kotlinOfficialDoc} and created a \microbench with 294 simple Kotlin programs and 135 simple Java programs, where each program demonstrates a single language construct. When considering taint analysis, we found that most Kotlin constructs can be analyzed the same way as the Java equivalents. However, we also found 18 engineering challenges that require a different approach. For example, functions declared as top-level elements do not have a parent class in the source code. However, the compiler generates a parent class in the Java bytecode, which the taint analysis should be aware of to locate the function correctly. We propose solutions for eight of these challenges that analysis writers can implement. As a proof of concept, we extended an existing Java taint analysis tool, \secucheck~\cite{secucheck}, by implementing six of our eight solutions, creating a taint analysis tool \secucheckkotlin that supports the standard language constructs. Finally, we evaluated the applicability of \secucheckkotlin with the Kotlin version of the PetClinic application\footnote{\url{https://github.com/spring-petclinic/spring-petclinic-kotlin}}.%\rk{confirm: contrast repo link or non-vulnerable repo link? GP: keep this link}

We present the details of our methodology in Section~\ref{sec:methodology}. Then, in Section~\ref{sec:findings}, we report on our findings from the study. Next, we present details of our implementation of \secucheckkotlin in Section~\ref{sec:secucheck-kotlin-tool}. Finally, we conclude and present our future work in Section~\ref{sec:conclusion}.
\section{Methodology}
\label{sec:methodology}

We examined the intermediate representation (IR) of the Kotlin code and---if existing---the equivalent Java code. Our methodology consists of automatic IR generation with metadata useful for our examination, which is a manual step that follows. We examined the following: (1) whether the generated IR for Kotlin is valid and can be analyzed the same way as the IR from equivalent Java code, (2) whether there are difficulties due to the definition of sources and sinks, and (3) whether there are language constructs in Kotlin that the analysis needs to handle in a new unique way when compared to Java. We did not consider challenges that can occur due to the callgraph-generation algorithms or computing alias information algorithms. 

We used Kotlin's official documentation~\cite{kotlinOfficialDoc} to examine each language construct. 
During the examination, we covered all constructs from the ``Concepts'' section and a few from the ``Standard library'' section (Collections, Iterators, Ranges, and Progressions). We did not consider constructs that were in the experimental stage at the time of this study. Table~\ref{table:overview-of-kotlin-constructs} summarizes Kotlin's constructs discussed in the official documentation and the those we manually examined. 

\begin{table}[!htbp]
	\scriptsize
	\begin{tabular}{ m{.31\textwidth} | >{\centering}m{.05\textwidth} | >{\centering}m{.005\textwidth}}
		
		\hline
		\multicolumn{1}{c|}{\textbf{Constructs}} & \textbf{\#Sub-constructs} & \multicolumn{1}{c}{\textbf{Supported}} \\
		\hline
		{Types, Control flow, Packages \& imports, Null safety, Equality, This expression, Destructuring declarations, Ranges and Progressions} & 11 & \multicolumn{1}{c}{\color{darkgreen}\ding{52}} \\
		\hline
		{Classes and objects (except for Delegated properties)} & 17 & \multicolumn{1}{c}{{\color{darkgreen}\ding{52}}} \\
		\hline
		{Functions (except for Builders)} & 5 & \multicolumn{1}{c}{{\color{darkgreen}\ding{52}}} \\
		\hline
		{Asynchronous programming techniques, Coroutines, Annotations, and Reflections} & 4 & \multicolumn{1}{c}{\color{red}\ding{56}} \\
		\hline
		{Collections and Iterators} & 2 & \multicolumn{1}{c}{\color{atomictangerine}\ding{52}} \\
		
		\hline
		
		\multicolumn{3}{c}{\scalebox{0.8}{Legend}} \\ [-1ex]
		\multicolumn{3}{c}{\scalebox{0.8}{{\color{darkgreen}\ding{52}}: examined all constructs in the category.}}\\ [-1ex]
		%\multicolumn{3}{c}{\scalebox{0.6}{{\color{darkgreen}\ding{52}}: examined all constructs in the category.}}\\ [-1ex]
		\multicolumn{3}{c}{\scalebox{0.8}{{\color{atomictangerine}\ding{52}}: examined only the basic constructs.}}\\ [-1ex]
		\multicolumn{3}{c}{\scalebox{0.8}{{\color{red}\ding{56}}: did not examine in this study.}} 
	\end{tabular}
	\caption{{\itshape List of Kotlin's features discussed in Kotlin's official documentation.}}
	\label{table:overview-of-kotlin-constructs}
%	\vspace{-6mm}
\end{table}	

Kotlin targets most Java Development Kit (JDK) versions. However, the annual developer ecosystem survey conducted by JetBrains in 2020 shows that 73\% of Kotlin developers target JDK 8~\cite{jetbrainsStatistics}. Furthermore, Kotlin targets JDK 8 by default. Therefore, we consider JDK 8 for this exploratory study. Additionally, we consider the Kotlin version 1.5.10. 

The Kotlin compiler has various options and annotations for modifying the compilation process, which alters the output of the compiler, Java bytecode. For this study, we used the default configuration of the compiler.

%\subsection{\jimplewrapper}
%\label{subsec:jimpleprovider}
%We consider Jimple \cite{jimple} for the manual examination because the Jimple code are more readable than the Java bytecode and represents the Java bytecode without losing any information. 
For the IR generation, we built a tool that generates Jimple IR using the Soot framework~\cite{soot}---\jimplewrapper. The Jimple code is organized based on the package name. Furthermore, for each class, \jimplewrapper generates metadata in a JSON file that contains information such as class name, super class, implemented interfaces, method count, method signatures, local variables, invoke expressions, etc. This metadata helps to identify the challenges easily and quickly. For deeper examination, we then examine the IR and Java bytecode.		
\subsection{\Cmicrobench}
\label{subsec:microbench}
Using real-world projects for the manual examination is infeasible because a real-world project has a complex mix of many constructs, making it hard to identify them clearly in Jimple. Therefore, we built a \microbench \ suite classified into two groups---Kotlin suite and Java suite. The Kotlin suite consists of small Kotlin programs, each focusing on one particular Kotlin construct. If a corresponding feature exists in Java, then an equivalent program is present in the Java suite. The suits contain six main categories: basics (43 Kotlin \& 36 Java files), classes and objects (118 Kotlin \& 80 Java files), functions (27 Kotlin \& 4 Java files), generics (8 Kotlin \& 10 Java files), unique to Kotlin (87 Kotlin files), and collection (11 Kotlin \& 5 Java files). Table \ref{table:overview-of-kotlin-suite} provides the overview of the Kotlin suite and the important features in the six categories.

\begin{table}[!htbp]
	\scalebox{0.78}{
		\small
		\begin{tabular}{ p{.13\textwidth} | >{\centering}m{.35\textwidth} | p{.059\textwidth}}
		
				\hline
				\textbf{Categories in \newline Kotlin suite} & \textbf{Major features} & \textbf{\#Kotlin \newline files} \\
				\hline
				{basics} & data types, control flow, package, import, exceptions, equality, operators, variables & 41  \\
				\hline
				{classesAndObjects} & classes, enum class, inline class, sealed class, nested / inner class, interface, functional interface (SAM), object expression, object declaration, delegation, qualified this, type aliases, visibility modifiers  & 118 \\
				\hline
				{functions} & simple functions, default arguments, local functions, infix notations. tail recursive function, varargs & 27 \\
				\hline
				{generics} & simple generic type, generic functions, raw types, upper bounds, & 8 \\
				\hline
				{uniqueToKotlin} & data class, destructuring declaration, extensions, higher-order functions, inline functions, null safety, operator overloading, primary constructor, properties, ranges, progressions, smart cast, string template, declaration site variance, type projection & 87 \\
				\hline
				{collection} & collection and iterators & 11 \\
		
				\hline
		
			\end{tabular}}
	\caption{{\itshape Overview of Kotlin suite.}}
	\label{table:overview-of-kotlin-suite}
%	\vspace{-6mm}
\end{table}

\subsection{Manual examination}
\label{subsec:setup-method}
The manual examination of the Jimple code was performed by the first author, who has more than 4.5 years of software development experience and is a Ph.D.\ student focusing on programming languages and static analysis. The more complex constructs, especially those specific to Kotlin or with differences from Java, were discussed with the second author, a Ph.D.\ student in the last year with expertise in the static analysis, and an external researcher with professional experience in Kotlin development. The examiners used the \jimplewrapper to generate the IR for the entire \microbench. Then, each construct was inspected manually. First, the generated metadata that provides information related to taint analysis is studied. Next, the generated IR is checked for a deeper examination. If more information is needed, then the generated bytecode is examined. Based on this, the examiner concluded whether a construct requires special handling in Kotlin taint analysis compared to Java taint analysis. 

\subsection{Threats to Validity}
\label{subsec:-research-threats}
Our study involves a manual step, making it possible that some of the findings are incomplete or incorrect. Furthermore, the programs written in the \microbench suite are based on personal experience. Therefore, some advanced use cases may be missing. As discussed earlier in this section, we considered the Kotlin version 1.5.10 and the target JDK 8. However, there is a risk that for some of the constructs, the Kotlin compiler may generate the bytecode differently for different versions. Also, for some constructs, the compiler may generate bytecode differently if some compiler options are used. As stated earlier, we only used the default configuration.%  \eb{I guess it can also depend on the compiler version, not just on the target version...}
\section{Findings}
\label{sec:findings}
In Sub-Section \ref{subsec:found_prob_with_proposed_sol}, we present the engineering challenges we identified and to which we have proposed a solution. Then, in Sub-Section \ref{subsec:found_prob_as_open_issues}, we present the engineering challenges, which we leave as open issues. Then, in Sub-Section \ref{subsec:research-questions}, we answer two research questions for the exploratory study.
\subsection{Engineering challenges with proposed solution}
\label{subsec:found_prob_with_proposed_sol}

\subsubsection{\textbf{Data type mapping}}
\label{subsubsec:data_type}
On the bytecode level, some data types in Kotlin are mapped to Java data types. For example, the non-nullable \keyword{kotlin.Int} is mapped to Java's \keyword{int}. Table \ref{tab:datatypesmapping} summarizes the data type mapping from Kotlin source code to the Java bytecode. Similarly, the compiler maps the function type to \keyword{kotlin.jvm.functions.Function*} in the Java bytecode as described in Table \ref{tab:functiontypesmapping}. This mapping is only affected by the number of parameters taken by the function type. The type of the parameters or return type will not affect the mapping. Note: the mapping described in Table \ref{tab:functiontypesmapping} is also valid for the respective nullable function types. Due to this data type mapping, the users must provide valid method signatures based on the Java bytecode to specify the source, sink, and other relevant method calls. However, it is cumbersome for the users to find the valid method signatures in big projects, making the tool not usable. \noindent

\begin{table}[!htbp]
	\scalebox{0.80}{
	\centering
	%\fontsize{12}{10}\selectfont
	\small
	\begin{tabular}{| p{.24\textwidth} | p{.32\textwidth} |}
		\hline
		\multicolumn{1}{| c |}{\textsc{Kotlin function type}} & \multicolumn{1}{ c |}{\textsc{Type in Java bytecode}} \\
		
		\hline
		Function type with 0 parameter, \newline e.g. \keyword{() $\rightarrow$ Int} & \keyword{kotlin.jvm.functions.Function0} \\
		\hline
		Function type with 1 parameter, \newline e.g. \keyword{(Byte) $\rightarrow$ Unit} & \keyword{kotlin.jvm.functions.Function1} \\
		\hline
		Function type with 2 parameters, \newline e.g. \keyword{(Int, Int) $\rightarrow$ Int} & \keyword{kotlin.jvm.functions.Function2} \\
		\hline
		\multicolumn{2}{| c |}{...} \\
		\hline
		Function type with 22 parameters & \keyword{kotlin.jvm.functions.Function22} \\
		\hline
		Function type with more than 22 \newline parameters & \keyword{kotlin.jvm.functions.FunctionN} \\
		\hline
	\end{tabular}}
	\caption{Kotlin function type mapping.}
	\label{tab:functiontypesmapping}
\end{table}

\begin{table*}[!htbp]
	\begin{subtable}[b]{.485\textwidth}
		\centering
		\fontsize{7}{9.5}\selectfont
		\begin{tabular}{| p{.45\textwidth} | p{.45\textwidth} |}
			\hline
			\textsc{Kotlin data type} & \textsc{Type in Java bytecode} \\
			
			\hline
			\rowcolor{lightgray}
			\multicolumn{2}{| c |}{\textsc{Special return types}} \\
			\hline
			\keyword{Nothing} & \keyword{java.lang.Void} \\
			\hline
			\keyword{Unit} & \keyword{void} \\
			
			\hline
			\rowcolor{lightgray}
			\multicolumn{2}{| c |}{\textsc{Basic types}} \\
			\hline
			\keyword{Byte} & \keyword{byte} \\
			\hline
			\keyword{Short} & \keyword{short} \\
			\hline
			\keyword{Int} & \keyword{int} \\
			\hline
			\keyword{Long} & \keyword{long} \\
			\hline
			\keyword{Char} & \keyword{char} \\
			\hline
			\keyword{Float} & \keyword{float} \\
			\hline
			\keyword{Double} & \keyword{double} \\
			\hline
			\keyword{Boolean} & \keyword{boolean} \\
			\hline
			
			\rowcolor{lightgray}
			\multicolumn{2}{| c |}{\textsc{Few built-in class}} \\
			\hline
			\keyword{Any} & \keyword{java.lang.Object} \\
			\hline
			\keyword{Cloneable} & \keyword{java.lang.Cloneable} \\
			\hline
			\keyword{Comparable} & \keyword{java.lang.Comparable} \\
			\hline
			\keyword{Enum} & \keyword{java.lang.Enum} \\
			\hline
			\keyword{Annotation} & \keyword{java.lang.Annotation} \\
			\hline
			\keyword{CharSequence} & \keyword{java.lang.CharSequence} \\
			\hline
			\keyword{String} & \keyword{java.lang.String} \\
			\hline
			\keyword{Number} & \keyword{java.lang.Number} \\
			\hline
			\keyword{Throwable} & \keyword{java.lang.Throwable} \\
			\hline
			
			\rowcolor{lightgray}
			\multicolumn{2}{| c |}{\textsc{Array types}} \\
			\hline
			\keyword{Array<Byte>} & \keyword{java.lang.Byte[]} \\
			\hline
			\keyword{Array<Short>} & \keyword{java.lang.Short[]} \\
			\hline
			\keyword{Array<Int>} & \keyword{java.lang.Integer[]} \\
			\hline
			\keyword{Array<Long>} & \keyword{java.lang.Long[]} \\
			\hline
			\keyword{Array<Char>} & \keyword{java.lang.Character[]} \\
			\hline
			\keyword{Array<Float>} & \keyword{java.lang.Float[]} \\
			\hline
			\keyword{Array<Double>} & \keyword{java.lang.Double[]} \\
			\hline
			\keyword{Array<Boolean>} & \keyword{java.lang.Boolean[]} \\
			\hline
			\keyword{Array<Any>} & \keyword{java.lang.Object[]} \\
			\hline
			\keyword{Array<*>} & \keyword{*[]} \\
			\hline
			
			\rowcolor{lightgray}
			\multicolumn{2}{| c |}{\textsc{Basic types Array}} \\
			\hline
			\keyword{ByteArray} & \keyword{byte[]} \\
			\hline
			\keyword{ShortArray} & \keyword{short[]} \\
			\hline
			\keyword{IntArray} & \keyword{int[]} \\
			\hline
			\keyword{LongArray} & \keyword{long[]} \\
			\hline
			\keyword{CharArray} & \keyword{char[]} \\
			\hline
			\keyword{FloatArray} & \keyword{float[]} \\
			\hline
			\keyword{DoubleArray} & \keyword{double[]} \\
			\hline
			\keyword{BooleanArray} & \keyword{boolean[]} \\
			\hline
			
			\rowcolor{lightgray}
			\multicolumn{2}{| c |}{\textsc{Immutable Collections}} \\
			\hline
			\keyword{Collection<T>} & \keyword{java.util.Collection<T>} \\
			\hline
			\keyword{List<T>} & \keyword{java.util.List<T>} \\
			\hline
			\keyword{Set<T>} & \keyword{java.util.Set<T>} \\
			\hline
			\keyword{Map<K, V>} & \keyword{java.util.Map<K, V>} \\
			\hline
			\keyword{Map.Entry<K, V>} & \keyword{java.util.Map.Entry<K, V>} \\
			\hline
			\keyword{Iterator<T>} & \keyword{java.util.Iterator<T>} \\
			\hline
			\keyword{Iterable<T>} & \keyword{java.lang.Iterable<T>} \\
			\hline
			\keyword{ListIterator<T>} & \keyword{java.util.ListIterator<T>} \\
			\hline
			
			\rowcolor{lightgray}
			\multicolumn{2}{| c |}{\textsc{Mutable Collections}} \\
			\hline
			\keyword{MutableCollection<T>} & \keyword{java.util.Collection<T>} \\
			\hline
			\keyword{MutableList<T>} & \keyword{java.util.List<T>} \\
			\hline
			\keyword{MutableSet<T>} & \keyword{java.util.Set<T>} \\
			\hline
			\keyword{MutableMap<K, V>} & \keyword{java.util.Map<K, V>} \\
			\hline
			\keyword{MutableMap.Entry<K, V>} & \keyword{java.util.Map.Entry<K, V>} \\
			\hline
			\keyword{MutableIterator<T>} & \keyword{java.util.Iterator<T>} \\
			\hline
			\keyword{MutableIterable<T>} & \keyword{java.lang.Iterable<T>} \\
			\hline
			\keyword{MutableListIterator<T>} & \keyword{java.util.ListIterator<T>} \\
			\hline
		\end{tabular}
		\caption{Mapping for non-nullable types}
		\label{subtab:nonnullablemapping}
	\end{subtable}%
	\hfil
	\begin{subtable}[b]{.485\textwidth}
		\centering
		\fontsize{7}{9.5}\selectfont
		\begin{tabular}{| p{.45\textwidth} | p{.45\textwidth} |}
			\hline
			\textsc{Kotlin data type} & \textsc{Type in Java bytecode} \\
			\hline
			
			\rowcolor{lightgray}
			\multicolumn{2}{| c |}{\textsc{Special return types}} \\
			\hline
			\keyword{Nothing?} & \keyword{java.lang.Void} \\
			\hline
			\keyword{Unit?} & \keyword{Unit} \\
			\hline
			
			\rowcolor{lightgray}
			\multicolumn{2}{| c |}{\textsc{Basic types}} \\
			\hline
			\keyword{Byte?} & \keyword{java.lang.Byte} \\
			\hline
			\keyword{Short?} & \keyword{java.lang.Short} \\
			\hline
			\keyword{Int?} & \keyword{java.lang.Integer} \\
			\hline
			\keyword{Long?} & \keyword{java.lang.Long} \\
			\hline
			\keyword{Char?} & \keyword{java.lang.Character} \\
			\hline
			\keyword{Float?} & \keyword{java.lang.Float} \\
			\hline
			\keyword{Double?} & \keyword{java.lang.Double} \\
			\hline
			\keyword{Boolean?} & \keyword{java.lang.Boolean} \\
			\hline
			
			\rowcolor{lightgray}
			\multicolumn{2}{| c |}{\textsc{Few built-in class}} \\
			\hline
			\keyword{Any?} & \keyword{java.lang.Object} \\
			\hline
			\keyword{Cloneable?} & \keyword{java.lang.Cloneable} \\
			\hline
			\keyword{Comparable?} & \keyword{java.lang.Comparable} \\
			\hline
			\keyword{Enum?} & \keyword{java.lang.Enum} \\
			\hline
			\keyword{Annotation?} & \keyword{java.lang.Annotation} \\
			\hline
			\keyword{CharSequence?} & \keyword{java.lang.CharSequence} \\
			\hline
			\keyword{String?} & \keyword{java.lang.String} \\
			\hline
			\keyword{Number?} & \keyword{java.lang.Number} \\
			\hline
			\keyword{Throwable?} & \keyword{java.lang.Throwable} \\
			\hline
			
			\rowcolor{lightgray}
			\multicolumn{2}{| c |}{\textsc{Array types}} \\
			\hline
			\keyword{Array<Byte>?} & \keyword{java.lang.Byte[]} \\
			\hline
			\keyword{Array<Short>?} & \keyword{java.lang.Short[]} \\
			\hline
			\keyword{Array<Int>?} & \keyword{java.lang.Integer[]} \\
			\hline
			\keyword{Array<Long>?} & \keyword{java.lang.Long[]} \\
			\hline
			\keyword{Array<Char>?} & \keyword{java.lang.Character[]} \\
			\hline
			\keyword{Array<Float>?} & \keyword{java.lang.Float[]} \\
			\hline
			\keyword{Array<Double>?} & \keyword{java.lang.Double[]} \\
			\hline
			\keyword{Array<Boolean>?} & \keyword{java.lang.Boolean[]} \\
			\hline
			\keyword{Array<Any>?} & \keyword{java.lang.Object[]} \\
			\hline
			\keyword{Array<*>?} & \keyword{*[]} \\
			\hline
			
			\rowcolor{lightgray}
			\multicolumn{2}{| c |}{\textsc{Basic types Array}} \\
			\hline
			\keyword{ByteArray?} & \keyword{byte[]} \\
			\hline
			\keyword{ShortArray?} & \keyword{short[]} \\
			\hline
			\keyword{IntArray?} & \keyword{int[]} \\
			\hline
			\keyword{LongArray?} & \keyword{long[]} \\
			\hline
			\keyword{CharArray?} & \keyword{char[]} \\
			\hline
			\keyword{FloatArray?} & \keyword{float[]} \\
			\hline
			\keyword{DoubleArray?} & \keyword{double[]} \\
			\hline
			\keyword{BooleanArray?} & \keyword{boolean[]} \\
			\hline
			
			\rowcolor{lightgray}
			\multicolumn{2}{| c |}{\textsc{Immutable Collections}} \\
			\hline
			\keyword{Collection<T>?} & \keyword{java.util.Collection<T>} \\
			\hline
			\keyword{List<T>?} & \keyword{java.util.List<T>} \\
			\hline
			\keyword{Set<T>?} & \keyword{java.util.Set<T>} \\
			\hline
			\keyword{Map<K, V>?} & \keyword{java.util.Map<K, V>} \\
			\hline
			\keyword{Map.Entry<K, V>?} & \keyword{java.util.Map.Entry<K, V>} \\
			\hline
			\keyword{Iterator<T>?} & \keyword{java.util.Iterator<T>} \\
			\hline
			\keyword{Iterable<T>?} & \keyword{java.lang.Iterable<T>} \\
			\hline
			\keyword{ListIterator<T>?} & \keyword{java.util.ListIterator<T>} \\
			\hline
			
			\rowcolor{lightgray}
			\multicolumn{2}{| c |}{\textsc{Mutable Collections}} \\
			\hline
			\keyword{MutableCollection<T>?} & \keyword{java.util.Collection<T>} \\
			\hline
			\keyword{MutableList<T>?} & \keyword{java.util.List<T>} \\
			\hline
			\keyword{MutableSet<T>?} & \keyword{java.util.Set<T>} \\
			\hline
			\keyword{MutableMap<K, V>?} & \keyword{java.util.Map<K, V>} \\
			\hline
			\keyword{MutableMap.Entry<K, V>?} & \keyword{java.util.Map.Entry<K, V>} \\
			\hline
			\keyword{MutableIterator<T>?} & \keyword{java.util.Iterator<T>} \\
			\hline
			\keyword{MutableIterable<T>?} & \keyword{java.lang.Iterable<T>} \\
			\hline
			\keyword{MutableListIterator<T>?} & \keyword{java.util.ListIterator<T>} \\
			\hline
		\end{tabular}
		\caption{Mapping for nullable types}
		\label{subtab:nullablemapping}
	\end{subtable}
	\caption{Data types mapping from Kotlin source code to the Java bytecode}
	\label{tab:datatypesmapping}
\end{table*}

%%%%%%%%%%%%%%%%%%%%%%%%%%%%%%%%%%%%%%%
%    Proposed solution
%%%%%%%%%%%%%%%%%%%%%%%%%%%%%%%%%%%%%%%
\textbf{Proposed solution: }
To handle this challenge, static-analysis developers can implement a data type transformer, which takes a method signature provided by the users as input. Then, the transformer checks for the parameters and return type in the given method signature. If the parameters type and return type are valid Kotlin data types, the transformer replaces the Kotlin data type with the respective Java data type.
%, as summarized in the official documentation %\footref{foot:datatype}. 

\subsubsection{\textbf{Type alias}}
%\customlabel{id:a2}{}
\label{subsubsec:type_alias}
A type alias allows developers to give a new name to the existing type. For example, in the Kotlin standard library, \keyword{ArrayList} is defined as a type alias to \keyword{java.util.ArrayList}. Therefore, \keyword{ArrayList} does not exist in the bytecode. However, the experts in the Kotlin programming language know which types are defined as type alias in Kotlin standard libraries. Furthermore, domain experts in custom libraries such as cryptographic APIs know what type aliases are defined in their libraries. On the other hand, users of the existing Java taint analysis tools may not know such type aliases and may give invalid method signatures.\\ \noindent
%%%%%%%%%%%%%%%%%%%%%%%%%%%%%%%%%%%%%%%
%    Proposed Solution
%%%%%%%%%%%%%%%%%%%%%%%%%%%%%%%%%%%%%%%
\textbf{Proposed solution: }
Static-analysis developers can implement a feature as part of the DSL that allows domain experts to specify type aliases---type alias specifications. The DSL semantics replaces all the type aliases found in the given method signatures with the original type specified in the given type alias specifications.
\subsubsection{\textbf{Property}}
%\customlabel{id:a3}{}
\label{subsubsec:property}
In Kotlin, a property is a field with an accessor. By default, Kotlin provides a getter and setter for mutable properties; for immutable properties, the getter only. Whenever there is access to a property in Kotlin source code, the Kotlin compiler uses the respective accessor method in the Java bytecode. Similar to variables, properties can be tainted. Therefore, the getter and setter of properties can be the source, sink, or propagator methods. Thus, the user needs to be aware of these signatures. 
%Suppose novice users of the existing Java taint analysis tools want to specify a getter or setter method of a property as a source or sink method. In that case, users must identify the valid method signature of the getter and setter methods. Therefore, static analysis developers must handle this problem in the DSL component to build the valid getter and setter method of a property.
\\ 
\noindent
%%%%%%%%%%%%%%%%%%%%%%%%%%%%%%%%%%%%%%%
%    Proposed Solution
%%%%%%%%%%%%%%%%%%%%%%%%%%%%%%%%%%%%%%%
\textbf{Proposed solution: }
Static-analysis developers can provide a feature in the DSL that enables users to specify a property by providing the fully qualified class name in which the property is defined, the property name, and the property's type. Then, the valid accessor method signature can be built automatically. The pattern for the getter method is \constant{<given fully qualified class name>: <given property's type> get<given property name with first letter caps>()}. Similarly, the setter method's pattern is \constant{<given fully qualified class name>: void set<given property name with first letter caps>(<given property's type>)}.
\subsubsection{\textbf{Top-level members}}
%\customlabel{id:a4}{}
\label{subsubsec:top-level-members}
In Kotlin, top-level members are defined in a Kotlin file under a package. Kotlin functions and properties can be top-level members. These members are not declared in any class, object, or interface. Therefore, in Kotlin source code, top-level members can be accessed directly without creating any object or using a class to access it. However, the Kotlin compiler generates a class in the Java bytecode and declares those top-level members as static members in the generated class. Suppose a novice user wants to specify top-level members as the source, sanitizer, propagator, or sink methods. In that case, the user must identify the valid class name in the method signature of top-level members.\\ \noindent
%%%%%%%%%%%%%%%%%%%%%%%%%%%%%%%%%%%%%%%
%    Proposed solution
%%%%%%%%%%%%%%%%%%%%%%%%%%%%%%%%%%%%%%%
\textbf{Proposed solution: }
To identify a valid class name of top-level members, one needs the filename and the package name in which top-level members are defined. Therefore, static-analysis developers can provide a feature in the DSL that enables users to specify a function or a property as a top-level member by providing the package name and the file name in which a top-level member is defined. Then, the DSL component can build a valid class name for a top-level function or accessors of a top-level property. The rule to build the valid class name is \constant{<given package name>.<given file name>Kt}.
\subsubsection{\textbf{Default arguments}}
%\customlabel{id:a5}{}
\label{subsubsec:default-arguments}
%In Java, the overloads feature can achieve a default value to function or constructor arguments. However, this increases the number of overloads. Kotlin avoids this problem by providing a default argument feature. In Kotlin, a constructor or function can have default arguments. For a function or constructor that have default arguments, the Kotlin compiler generates two implementations in the Java bytecode. First implementation is the actual implementation of a function or constructor with all the parameters defined in Kotlin source code. The second implementation contains additional arguments that determines which arguments default value to consider and then calls the actual implementation. 
%If the developer does not pass value to default arguments, then the Kotlin compiler calls the second implementation. For a default argument in a function, the Kotlin compiler adds the suffix \constant{\$default} to the function name for the second implementation. Suppose users of taint analysis tools specify a constructor or function that has default argument as a source or propagator method. In that case, the analysis component should identify the second implementation generated by the Kotlin compiler as a source or propagator method and track the variables correctly. %Otherwise, the existing Java taint analysis tools fail to detect taint-flows in default argument feature.
In Java, the overload feature can achieve a default value to function or constructor arguments. However, this increases the number of overloads. Kotlin avoids this problem by providing a default argument feature in a constructor or function. For a function or constructor with default arguments, the Kotlin compiler generates two implementations in the Java bytecode. First, the actual implementation with all the parameters as defined in source code. Second implementation generated by the compiler with additional arguments that determines the default arguments' value and calls the actual implementation. For constructor, the compiler adds two additional arguments at the end---\keyword{int} and \keyword{kotlin.jvm.internal.DefaultConstructorMar-\\ker}. Similarly, for a function, the compiler adds \keyword{int} and \keyword{java.lang.Object} at the end. Additionally, if the function is a member function, the compiler adds a first argument of type in which the function is defined. This added first argument is the \keyword{this}-object of the member function. Furthermore, for a default argument in a top-level function or member function, the compiler adds the suffix \constant{\$default} to the function name for the second implementation. If a developer does not pass value to default arguments, then the compiler calls the second implementation. Suppose users of taint analysis tools specify a default argument constructor or function as a source method. In that case, the analysis component should identify the second implementation generated by the compiler as a source method and track the variables correctly. %Otherwise, the existing Java taint analysis tools fail to detect taint-flows in default argument feature.
\\ \noindent
%%%%%%%%%%%%%%%%%%%%%%%%%%%%%%%%%%%%%%%
%    Proposed solution
%%%%%%%%%%%%%%%%%%%%%%%%%%%%%%%%%%%%%%%
\textbf{Proposed solution: }
If the analysis fails to identify a method call as a source, sink, or other relevant method as specified in taint-flow specifications, then the analysis checks for the second implementation of the default argument feature. For each function or constructor in taint-flow specifications, add the additional arguments and modify the function name as described in Sub-Section \ref{subsubsec:default-arguments}. Subsequently, if the method signature matches with the method call's signature, track the respective variables. For constructor and top-level function, track the variables based on the specified rules for the matched method in taint-flow specifications. However, for member functions, since the compiler adds a parameter at the beginning, the analysis should consider this added first argument while tracking the variables. For example, if the \keyword{this}-object is specified to track, then track the first argument in the Java bytecode. Likewise, track the second argument in the Java bytecode if the first argument is specified to track and so forth.

\subsubsection{\textbf{Extensions}}
%\customlabel{id:a6}{}
\label{subsubsec:extensions}
In Kotlin, the extension feature allows extending an existing class with new members without using inheritance. However, extensions will not modify and add a new member to an existing class; instead, the new member is made accessible using the dot-notation on variables of the type (receiver type) for which the extension member is defined.  In the Java bytecode for a top-level extension function, the Kotlin compiler adds the receiver type as the first argument, followed by the actual parameters defined in the source code. Similarly, the compiler adds the receiver type as the first argument to the getter method of a top-level extension property. Note: The compiler generates only the getter method for an extension property. Furthermore, for top-level companion object extension members, the compiler also adds the receiver type as the first argument, followed by the actual argument defined in the source code. However, the added first argument type is the wrapper class generated for a companion object. The companion object is discussed in detail in Section~\ref{subsubsec:companion-object-open-prob}. Like top-level extension members, the compiler also adds the receiver type as the first argument for an extension defined as a class member. Furthermore, Kotlin supports qualified \keyword{this}-object to access the outer class's \keyword{this}-object. For this, the compiler considers the actual \keyword{this}-object (outer class's \keyword{this}) in the Java bytecode as a qualified outer class's \keyword{this}-object in the source code and the first argument in the Java bytecode as a receiver \keyword{this}-object in the source code. Suppose users want to specify an extension member as a source or sink method, then users might give an invalid method signature since users might not be aware of the first argument of receiver type added by the compiler. Furthermore, if users specify to track the \keyword{this}-object in an extension member, then the analysis should track the first argument. Likewise, the analysis should track the actual \keyword{this}-object in the Java bytecode if the outer class \keyword{this}-object is specified to track. Similarly, if users specify to track the first argument in an extension function, then the analysis should track the second argument and so forth. \\ \noindent
%%%%%%%%%%%%%%%%%%%%%%%%%%%%%%%%%%%%%%%%%%%%%%
%     Proposed solution
%%%%%%%%%%%%%%%%%%%%%%%%%%%%%%%%%%%%%%%%%%%%%%
\textbf{Proposed solution: }
To handle extension functions and extension properties, static-analysis developers should make their taint-flow specifiations aware of these. %can provide a feature in the DSL component that enables the users to specify a function or property as an extension member by providing the fully qualified class name for which the extension is defined. 
If this is done through the DSL for taint-flow specifications, the DSL can build the valid method signature by adding the given fully qualified class name as the first argument. Furthermore, the users should not be able to obtain a setter method from an extension property since an extension property can not have a setter method. 

To handle companion object extensions, static-analysis developers can provide a feature in the DSL. This feature enables the users to specify a function or property as a companion object extension member by providing the fully qualified class name and the name of the companion object for which the extension is defined. If the name of the companion object is not given, then by default, the name is \constant{Companion}. From these inputs, the generated wrapper class for the companion object can be built as \constant{<given fully qualified class name>\$<given companion object name>}. Then, the valid method signature can be built by adding this wrapper class as a first argument. 

%%%%%%%%%%%%%%%%%%%%%%%%%%%%%%%%%%%%%%%%%%%%%%%%%%%%%%%%%%%%%%%%%%%%%%%%%%%%%%%%%%%%
%%
%%     This below table is for Operator Overloading. It is placed here so that it appears on the same page as operator overloading section
%%
%%%%%%%%%%%%%%%%%%%%%%%%%%%%%%%%%%%%%%%%%%%%%%%%%%%%%%%%%%%%%%%%%%%%%%%%%%%%%%%%%%%%
\begin{table*}[!htbp]
	\begin{subtable}[b]{.485\textwidth}
		\centering
		\fontsize{7}{9.5}\selectfont
		\begin{tabular}{| p{.45\textwidth} | p{.45\textwidth} |}
			\hline
			\textsc{Built-in operator} & \textsc{Mapped to a function} \\
			
			\hline
			\rowcolor{lightgray}
			\multicolumn{2}{| c |}{\textsc{Unary operators}} \\
			\hline
			\texttt{+obj} & \texttt{obj.unaryPlus()} \\
			\hline
			\texttt{-obj} & \texttt{obj.unaryMinus()} \\
			\hline
			\texttt{!obj} & \texttt{obj.not()} \\
			\hline
			\texttt{++obj} & \texttt{obj.inc()} \\
			\hline
			\texttt{-{}-obj} & \texttt{obj.dec()} \\
			\hline
			\texttt{obj++} & \texttt{obj.inc()} \\
			\hline
			\texttt{obj-{}-} & \texttt{obj.dec()} \\
			
			\hline
			\rowcolor{lightgray}
			\multicolumn{2}{| c |}{\textsc{Arithmetic operators}} \\
			\hline
			\texttt{obj1 + obj2} & \texttt{obj.plus(obj2)} \\
			\hline
			\texttt{obj1 - obj2} & \texttt{obj.minus(obj2)} \\
			\hline
			\texttt{obj1 * obj2} & \texttt{obj.times(obj2)} \\
			\hline
			\texttt{obj1 / obj2} & \texttt{obj.div(obj2)} \\
			\hline
			\texttt{obj1 \% obj2} & \texttt{obj.rem(obj2)} \\
			\hline
			\texttt{obj1..obj2} & \texttt{obj.rangeTo(obj2)} \\
			\hline
			
			\rowcolor{lightgray}
			\multicolumn{2}{| c |}{\textsc{Augmented assignment operators}} \\
			\hline
			\texttt{obj1 += obj2} & \texttt{obj.plusAssign(obj2)} \\
			\hline
			\texttt{obj1 -= obj2} & \texttt{obj.minusAssign(obj2)} \\
			\hline
			\texttt{obj1 *= obj2} & \texttt{obj.timesAssign(obj2)} \\
			\hline
			\texttt{obj1 /= obj2} & \texttt{obj.divAssign(obj2)} \\
			\hline
			\texttt{obj1 \%= obj2} & \texttt{obj.remAssign(obj2)} \\
			
			\hline
			\rowcolor{lightgray}
			\multicolumn{2}{| c |}{\textsc{Equality check operator}} \\
			\hline
			\texttt{obj1 == obj2} & \texttt{obj.equals(obj2)} \\
			\hline
			\texttt{obj1 != obj2} & \texttt{!(obj.equals(obj2))} \\
			\hline
		\end{tabular}
	\end{subtable}%
	\hfil
	\begin{subtable}[b]{.485\textwidth}
		\centering
		\fontsize{7}{9.5}\selectfont
		\begin{tabular}{| p{.40\textwidth} | p{.50\textwidth} |}
			\hline
			\textsc{Built-in operator} & \textsc{Mapped to a function} \\
			
			\hline
			\rowcolor{lightgray}
			\multicolumn{2}{| c |}{\textsc{In operator}} \\
			\hline
			\texttt{obj1 in obj2} & \texttt{obj.contains(obj2)} \\
			\hline
			\texttt{obj1 !in obj2} & \texttt{!(obj.contains(obj2))} \\
			
			\hline
			\rowcolor{lightgray}
			\multicolumn{2}{| c |}{\textsc{Index operators}} \\
			\hline
			\texttt{obj[i]} & \texttt{obj.get(i)} \\
			\hline
			\texttt{obj[i, j]} & \texttt{obj.get(i, j)} \\
			\hline
			\texttt{obj[i, j, k]} & \texttt{obj.get(i, j, k)} \\
			\hline
			\texttt{obj[i1, ..., in]} & \texttt{obj.get(i1, ..., in)} \\
			\hline
			\texttt{obj[i] = obj2} & \texttt{obj.set(i, obj2)} \\
			\hline
			\texttt{obj[i, j] = obj2} & \texttt{obj.set(i, j, obj2)} \\
			\hline
			\texttt{obj[i, j, k] = obj2} & \texttt{obj.set(i, j, k, obj2)} \\
			\hline
			\texttt{obj[i1, ..., in] = obj2} & \texttt{obj.set(i1, ..., in, obj2)} \\
			\hline
			
			\rowcolor{lightgray}
			\multicolumn{2}{| c |}{\textsc{Invoke operators}} \\
			\hline
			\texttt{obj()} & \texttt{obj.invoke()} \\
			\hline
			\texttt{obj(i)} & \texttt{obj.invoke(i)} \\
			\hline
			\texttt{obj(i, j)} & \texttt{obj.invoke(i, j)} \\
			\hline
			\texttt{obj(i, j, k)} & \texttt{obj.invoke(i, j, k)} \\
			\hline
			\texttt{obj(i1, i2, ..., in)} & \texttt{obj.invoke(i1, i2, ..., in)} \\
			\hline
			
			\rowcolor{lightgray}
			\multicolumn{2}{| c |}{\textsc{Comparison operators}} \\
			\hline
			\texttt{obj1 > obj2} & \texttt{obj.compareTo(obj2)} \\
			\hline
			\texttt{obj1 < obj2} & \texttt{obj.compareTo(obj2)} \\
			\hline
			\texttt{obj1 >= obj2} & \texttt{obj.compareTo(obj2)} \\
			\hline
			\texttt{obj1 <= obj2} & \texttt{obj.compareTo(obj2)} \\
			\hline
			& \\
			\hline
		\end{tabular}
	\end{subtable}
	\caption{Built-in operators and its corresponding functions in Kotlin.}
	\label{tab:operatorFunctionMappin}
\end{table*}

To handle the qualified \keyword{this}-object in extensions as members, the DSL should be able to track the \keyword{this}-object as extension receiver or dispatch receiver (outer class's \keyword{this}-object). If users specify to track \keyword{this}-object as an extension receiver, modify the taint-flow specification to track the first parameter in the Java bytecode. Similarly, if users specify to track \keyword{this}-object as dispatch receiver, modify the taint-flow specification to track the actual \keyword{this}-object in the Java bytecode. Similarly, for an extension function, if user specify to track the first parameter, then analysis should track the second parameter and so forth.
\subsubsection{\textbf{Infix function}}
%\customlabel{id:a7}{}
\label{subsubsec:infix-function}
In Kotlin, infix functions are called using the infix notation, i.e., without the dot notation and the parentheses. The infix function must be a member function or extension function and must have a single parameter without a default value. Similar to a standard function, an infix function can be a source, sink, and other relevant methods. However, a novice user of taint analysis tools may not know how the infix function works in the Java bytecode and may provide invalid method signatures.\\ \noindent
%%%%%%%%%%%%%%%%%%%%%%%%%%%%%%%%%%%%%%%%%%%%%%
%     Proposed solution
%%%%%%%%%%%%%%%%%%%%%%%%%%%%%%%%%%%%%%%%%%%%%%
\textbf{Proposed solution: }
Static-analysis developers can provide a DSL feature that enables users of taint analysis tools to specify a function as an infix function by providing a function name, receiver type, parameter type, and return type. Then, DSL can build a valid method signature as \constant{<given receiver type>: <given return type> <given function name>(<given parameter type>)}. 
\subsubsection{\textbf{Operator overloading}}
%\customlabel{id:a8}{}
\label{subsubsec:operator-overloading}
Operator overloading redefines the implementation of the built-in operators with specific types. For example, one can overload the \texttt{++} operator by defining the function \funRef{inc} on a custom class. The compiler calls the implemented \funRef{inc} function in the Java bytecode. Table \ref{tab:operatorFunctionMappin} provides the mapping between the built-in operator and the function name. An overloaded operator function can be a sanitizer or propagator method. However, the novice users of taint analysis tools may not know the mapping of the built-in operators to the function name and may provide invalid method signatures. 
\\ \noindent
%%%%%%%%%%%%%%%%%%%%%%%%%%%%%%%%%%%%%%%%%%%%
%     Proposed solution
%%%%%%%%%%%%%%%%%%%%%%%%%%%%%%%%%%%%%%%%%%%%
\textbf{Proposed solution: }
Static-analysis developers can provide a feature in DSL that enables users to specify an overloaded operator by providing the symbol of an operator, type of the receiver, return type, and the parameter(s) type based on an operator. Then, DSL can build the valid method signature by mapping the given operator symbol to the function as described in Table \ref{tab:operatorFunctionMappin}.
\subsection{Engineering challenges without solution (open issues)}
\label{subsec:found_prob_as_open_issues}

\subsubsection{\textbf{Companion object}}
%\customlabel{id:b1}{}
\label{subsubsec:companion-object-open-prob}
In Kotlin, a companion object binds members to a class rather than the instance of a class. Kotlin's companion object is similar to Java's static members. However, the Kotlin compiler generates a wrapper class for each companion object in the Java bytecode. The naming scheme for that wrapper class is \constant{<class name in which the companion object is defined>\$<companion object name>}. If the companion object name is not provided in Kotlin source code, then by default the name is \constant{Companion}. The compiler places the implementation of that companion object's members in the generated wrapper class. 

Furthermore, to allow that wrapper class to access the private members of the actual class and vice versa, the compiler generates additional functions for each private member. For a private function, the naming scheme for the generated function is \constant{access\$<actual name of the private function>}. Similarly, the naming scheme for the accessors of a private property is \constant{access\$<accessor's method name of a property>\$cp}. The accessors' method name is discussed in Sub-Section \ref{subsubsec:property}.

Due to such implementation of companion objects in the Java bytecode, users of taint analysis tools might find it difficult to identify valid method signatures. Additionally, for the function that takes a companion object as a parameter, users must give that parameter type as a generated wrapper class in the method signature, which is not visible in the source code. Furthermore, the analysis should be aware of the generated functions for private members, which might be a possible source, sink, or propagator.
\subsubsection{\textbf{Destructuring declaration}}
%\customlabel{id:b2}{}
\label{subsubsec:destructuring-dec-open-prob}
In Kotlin, an object can be destructured into multiple variables in a single statement using the destructuring declaration. To allow a class to destructure, that class must have the \funRef{component\textit{N}} functions with the \keyword{operator} keyword. These component functions return the properties of a class. The widely used convention for the order of \funRef{component\textit{N}} functions is the order of properties defined in a class. However, it is not mandatory, and developers can make component functions return any properties of a class. Suppose the function \funRef{component1} returns the first property and the users of taint analysis tools specify the getter method of the first property as a source method. In that case, the analysis should be able to identify the \funRef{component1} function as a source method. Therefore, the analysis must know the mapping between the \funRef{component\textit{N}} functions and properties of a class to identify a taint-flow in a destructuring declaration. 
\subsubsection{\textbf{Internal modifier}}
%\customlabel{id:b3}{}
\label{subsubsec:internal-modifier-open-prob}
In Kotlin, a member declared with an \keyword{internal} modifier is only visible inside the module in which the member is defined. Kotlin defines a module as a group of Kotlin files that are compiled together. In the Java bytecode, the Kotlin compiler appends the symbol hyphen followed by the module name for the accessors of an internal property and to an internal member function. However, we did not observe this behavior for classes, interfaces, top-level functions, or accessors of top-level properties, which are declared as \keyword{internal}. Suppose users of taint analysis tools specify an internal member function or accessors of internal property as a sink method. In that case, the analysis component should identify the modified name with the appended module name as a sink method. Otherwise, the analysis component fails to detect taint-flow in \keyword{internal} member functions and properties. Note: if there is a symbol hyphen in the module name, the Kotlin compiler replaces it with the underscore before appending it to the internal member functions and accessors of internal property in the Java bytecode.
\subsubsection{\textbf{Inline class}}
%\customlabel{id:b4}{}
\label{subsubsec:inline-class-open-prob}
Kotlin's inline class wraps an existing class with improved performance compared to a manually created wrapper class. In the Java bytecode, the Kotlin compiler generates some of the member functions for an inline class---constructor, accessor for a property (wrapped class), \funRef{toString}, \funRef{hashCode}, and equality check. These functions are generated to support the interoperability with Java. However, the compiler generates the alternative version of these functions to improve the performance by inlining the wrapped class in place of wrapper class usage. In addition, the compiler adds the suffix \constant{-impl} to the improved version of these functions and to the overridden function of an interface. Additionally, the compiler generates \funRef{box-impl} and \funRef{unbox-impl} function for boxing and unboxing the wrapped class. The Kotlin compiler calls the \funRef{-impl} version of member functions wherever it is possible to improve the performance. Suppose users of taint analysis tools specify the member functions of an inline class as a source. In that case, along with the actual implementation, the analysis should identify its \constant{-impl} version as a source. Otherwise, the existing Java taint analysis tools fail to detect taint-flows in an inline class.
\subsubsection{\textbf{Function returning anonymous object}}
%\customlabel{id:b5}{}
\label{subsubsec:fun-ret-anonymous-obj-open-prob}
In Kotlin, object expressions create objects of an anonymous class. Every object expression has at least one base class. The Kotlin compiler generates a wrapper class for each instance of object expression in the Java bytecode similar to Java. However, in contrast to Java, the return type in Kotlin's function is not mandatory to specify, and the compiler can infer the type. Suppose a function is private and returns an anonymous object. In that case, the compiler infers the return type as the generated wrapper class, which is not visible in the source code. This makes it challenging for the users to identify the valid method signature of a private function that returns anonymous object.
\subsubsection{\textbf{Local functions}}
%\customlabel{id:b6}{}
\label{subsubsec:local-functions-open-prob}
Kotlin supports local functions, which are functions inside other functions. These local functions can access the outer functions local variables. For a local function, the Kotlin compiler generates a static function in the Java bytecode. The naming scheme and the parameters of this static function are \constant{<outer function name>\$<local function name><-digits starting from 0 if there are multiple local function with th-\\e same name>(<outer functions local varia-\\bles if accessed by local function>, <\keyword{this} object if the outer function is a member function>, <actual parameter as defined f-\\or the local function in Kotlin source co-\\de>)}. Additionally, if a local function accesses an mutable local variable of an outer function, then the compiler passes the reference type to reflect the changes in the outer function. For example, if the local function access mutable \keyword{Int} type, then in the Java bytecode the Kotlin compiler passes the \keyword{kotlin.jvm.internal.Ref\$IntRef} type to the generated static function as a parameter. Due to such implementation of local function in the Java bytecode, it is challenging for the user to identify the valid method signature of a local function. Furthermore, the analysis must handle the accessed local variables of the outer functions to track the tainted variable. 
\subsubsection{\textbf{Higher-order functions}}
%\customlabel{id:b7}{}
\label{subsubsec:higher-order-function-open-prob}
Kotlin provides a function type that enables higher-order function in Kotlin. These function types are mapped to \keyword{kotlin.jvm.functions.Functi-\\on*} types in the Java bytecode as described in Table \ref{tab:functiontypesmapping}. Furthermore, there are five ways to create an instance of a function type in Kotlin---lambda expression, anonymous function, function literal with a receiver, callable reference, and instances of a custom class that implements a function type. The Kotlin compiler generates a wrapper class for each instance of a function type in Kotlin source code. The naming scheme for this wrapper class is \constant{<class name in which the lambda is declared>\$<function name in which the lambda is declared>\$<variable n-\\ame in which the lambda expression is sto-\\red if any otherwise this is optional>\$<d-\\igits starting from 1>}. This wrapper class overrides the interface function \funRef{invoke}, in which the Kotlin compiler places the implementation of a lambda expression.

Similar to the local functions accessing the outer function's local variables as discussed in Sub-Section \ref{subsubsec:local-functions-open-prob}, lambda expressions can also access the outer function's local variables. All the accessed variables are passed to the constructor of the wrapper class. Then, the constructor stores these values in its fields, which can be accessed in the \funRef{invoke} method. Furthermore, if the outer function's local variable is immutable, the compiler passes the reference type, e.g. \keyword{kotlin.jvm.internal.Ref\$IntRef}. For an anonymous function, the compiler generates the Java bytecode similar to the lambda expression. Similarly, for a class implementing a function type, the compiler implements the \keyword{kotlin.jvm.functions.Function*} in the Java bytecode and implements the interface method \funRef{invoke}. 

For a function literal with a receiver, the compiler generates the Java bytecode similar to the lambda expression, except that the receiver object is passed as the first argument to the \funRef{invoke} method. For callable reference, the compiler generates the Java bytecode similar to a lambda expression. However, the receiver of a callable reference is passed to the constructor of the generated wrapper class, which stores the receiver in the superclass' field called \textit{receiver}. Later, the function \funRef{invoke} access the field \textit{receiver} to call the respective member function.

Java uses \keyword{invokedynamic} instruction for lambda expression. Therefore, the existing Java taint analysis tools detect taint-flows in lambda expressions in Java by handling the \keyword{invokedynamic} instruction in the Java bytecode. However, by default, the Kotlin compiler does not use \keyword{invokedynamic} instruction for an instance of a function type, which leads to the existing Java taint analysis tools failing to detect taint-flows in higher-order functions. Therefore, the analysis must handle the generated wrapper class for an instance of a function type to track the tainted information. Furthermore, the analysis should handle the \textit{receiver} property to track the tainted receiver object for a callable reference. Furthermore, similar to local functions (\ref{subsubsec:local-functions-open-prob}), the analysis should handle the accessed local variables of the outer functions to track the tainted variable. 

Note:  for a functional interface or a Single Abstract Method (SAM), the Kotlin compiler generates the Java bytecode similar to the Java's lambda expression by default, i.e., \keyword{invokedynamic} instruction in the Java bytecode.
\subsubsection{\textbf{Inline function}}
%\customlabel{id:b8}{}
\label{subsubsec:inline-fun-open-prob}
As discussed in Sub-Section \ref{subsubsec:higher-order-function-open-prob}, the Kotlin compiler generates a wrapper class for each instance of a function type, captures the outer function's local variables, which leads to extra memory allocations, and extra virtual method call introduces runtime overhead. However, in some scenarios, such runtime overhead can be eliminated by inlining the lambda expression rather than creating an instance of a function type. For this purpose, Kotlin provides inline functions. For example, the \funRef{println} function in Kotlin is declared as inline, which calls the Java's function \funRef{System.out.println}. Therefore, in the Java bytecode, we find the \funRef{System.out.println} function call in place of Kotlin's \funRef{println} call site. Similarly, custom higher-order functions can also be declared as inline in Kotlin. Suppose users of taint analysis tools specify an inline function as a sink method. In that case, taint analysis tools fail to detect taint-flow that reaches this sink method since there is no actual method call of an inline function in the Java bytecode. Therefore, taint analysis tools must know the propagation rule for all the method calls in the body of that inline function. Otherwise, it fails to detect taint-flows in inline functions.
\subsubsection{\textbf{Sealed class}}
%\customlabel{id:b9}{}
\label{subsubsec:sealed-class-open-prob}
A sealed class restricts users from inheriting a class or interface, and all the derived classes are known at compile time. To achieve this, the Kotlin compiler makes the constructor private and overloads the constructor with an additional parameter at the end---\keyword{kotlin.jvm.internal.DefaultConsructorMark-\\er}. This allows the compiler to call the overloaded constructor for the known derived class and restricts developers from creating a new derived class. Suppose users of taint analysis tools specify the constructor of a sealed class as a propagator method. In that case, the analysis must identify the overloaded constructor as a propagator. Otherwise, taint analysis tools fail to detect taint-flows in a sealed class's constructor.
\subsubsection{\textbf{Package}}
%\customlabel{id:b10}{}
\label{subsubsec:package-open-prob}
In Java, the package name must match the path of that Java file. However, in Kotlin, the package name can be different than the path of that Kotlin file. Once the analysis component completes and returns the found results, some existing Java taint analysis tools use the package name to build the path of the Java file to display the errors in an IDE. However, if the Kotlin file's path is different from its package, then taint analysis tools fail to display the found taint-flows in an IDE. 
\subsection{Research Questions}
\label{subsec:research-questions}
In the previous two sub-sections, Sub-Section \ref{subsec:found_prob_with_proposed_sol} and Sub-Section \ref{subsec:found_prob_as_open_issues}, we discussed the various engineering challenges that must be handled in the existing Java taint analysis tools to support taint analysis on Kotlin code. In this sub-section, we answer two research questions (\hyperref[rqs:rq1]{\RQ{1}} and \hyperref[rqs:rq2]{\RQ{2}}), which evaluates our exploratory study. \\[0.1em]
\customlabel{rqs:rq1}{}
\noindent
\RQ{1:} \textit{Which Kotlin's features can be analyzed by the existing Java taint analysis tools without any engineering challenge?}

To answer this research question, we list the Kotlin's features for which the Kotlin compiler generates the Java bytecode similarly to the Java compiler. The existing Java taint analysis tools can analyze Kotlin programs containing these features without any engineering challenges. For all the features listed under this research question, Soot generates the valid Jimple code.  Furthermore, the analysis component can perform taint analysis on these features and requires no additional constructs in the DSL component to handle these features.

\begin{table*}[!htbp]
	\begin{adjustbox}{max width=2\columnwidth}
		\renewcommand{\arraystretch}{1}% Wider
		\Huge
		\begin{tabular}{ m{1.5\textwidth} | >{\centering}m{.20\textwidth} | >{\centering\arraybackslash}m{\textwidth}}
			
			\hline
			\multicolumn{1}{c|}{\textbf{Kotlin's features}} & \textbf{Similarity level} & \textbf{similar to} \\
			
			\hline
			{Explicit conversion} & {\color{darkgreen}\ding{52}}* & {typecasting and Java's methods like \funRef{intValue}, \funRef{byteValue} etc.}  \\
			
			\hline
			{Arithmetic operators, Bitwise operators, Comparison operators, assignment operators, unary operators, logical operators, equality check, Literal constants, varargs} & {\color{darkgreen}\ding{52}} &  \\
			
			\hline
			{is operator, unsafe cast operator, safe cast operator} & {\color{darkgreen}\ding{52}}* & instance check (\keyword{instanceof}), typecasting  \\
			
			\hline
			{when construct} & {\color{darkgreen}\ding{52}}* & \keyword{lookupswitch}, \keyword{tableswitch}, comparison, goto and label  \\
			
			\hline
			{for construct} & {\color{darkgreen}\ding{52}}* & Java's for, iterators \\
			
			\hline
			{while, do-while, if construct} & {\color{darkgreen}\ding{52}} &  \\
			
			\hline
			{return, break, continue, labeled break and labeled continue, and qualified \keyword{this} in nested / inner class} & {\color{darkgreen}\ding{52}} & \\
			
			\hline
			{labeled return (non-local return)} & {\color{darkgreen}\ding{52}}* & goto and label statements \\
			
			\hline
			{try-catch, finally, throw} & {\color{darkgreen}\ding{52}} &  \\
			
			\hline
			{import, named arguments} & {\color{darkgreen}\ding{52}} & \\
			
			\hline
			{Open class} & {\color{darkgreen}\ding{52}} & non-final class \\
			
			\hline
			{Abstract class, inheritance, overriding methods, calling super class implementations, multiple inheritance} & {\color{darkgreen}\ding{52}} & \\
			
			\hline
			{Functional interface (SAM)} & {\color{darkgreen}\ding{52}}* & Java's lambda expression (invokedynamic) \\
			
			\hline
			{Generics} & {\color{darkgreen}\ding{52}} & \\
			
			\hline
			{Nested class, inner class, enum class} & {\color{atomictangerine}\ding{52}} & \\
			
			\hline
			{Object expression} & {\color{atomictangerine}\ding{52}} & instance of anonymous class \\
			
			\hline
			{Object declaration} & {\color{darkgreen}\ding{52}}* & singleton pattern \\
			
			\hline
			{Delegation (in inheritance)} & {\color{darkgreen}\ding{52}}* & Delegation pattern \\
			
			\hline
			{varargs} & {\color{darkgreen}\ding{52}} &  \\
			
			\hline
			{Tail-recursive function} & {\color{darkgreen}\ding{52}} & normal function and loops \\
			
			\hline
			{String template} & {\color{darkgreen}\ding{52}}* & Java's StringBuilder \funRef{append}, Kotlin's \funRef{stringPlus} methods \\
			
			\hline
			{Smart cast} & {\color{darkgreen}\ding{52}}* & Typecasting after instance check \\
			
			\hline
			{lateinit} & {\color{darkgreen}\ding{52}}* & uninitialized field, null check, Null pointer exception (NPE) \\
			
			\hline
			{Null safety} & {\color{darkgreen}\ding{52}}* & null check, goto and label statements, NPE  \\
			
			\hline
			{Default implementation in interface} & {\color{red}\ding{52}} &  \\
			
			\hline
			
			\multicolumn{3}{c}{} \\
			\multicolumn{3}{c}{Legend} \\
			\multicolumn{3}{c}{{\color{darkgreen}\ding{52}}: similar to the respective feature in Java.}\\
			\multicolumn{3}{c}{{\color{atomictangerine}\ding{52}}: similar to Java, but the naming scheme of the generated wrapper class is different compared to Java.} \\
			\multicolumn{3}{c}{{\color{red}\ding{52}}: completely different to Java, but there is no challenge concerning taint analysis in DSL, analysis or IR generator components.} \\
			\multicolumn{3}{c}{*: similar to Java's features, some of them are not visible in the source code.} \\
		\end{tabular}
		
	\end{adjustbox}
	\caption{List of Kotlin's features, for which the existing Java taint analysis tools can analyze without any challenge.}
	\label{table:evaluation-no-problem-table}
	
\end{table*}

\begin{table*}[!htbp]
	\begin{adjustbox}{max width=2\columnwidth}
		\renewcommand{\arraystretch}{1}% Wider
		\begin{tabular}{ m{.40\textwidth} | >{\centering}m{.40\textwidth} | >{\centering}m{.10\textwidth} | m{.35\textwidth}}
			
			\hline
			\multicolumn{1}{c|}{\textbf{Kotlin's features}} & \textbf{Engineering challenges} & \textbf{can be solved in} & \multicolumn{1}{c}{\textbf{Note}} \\
			
			\hline
			{Data types} & Data type mapping (\ref{subsubsec:data_type}) & {\color{blue}\ding{108}} &   \\
			
			\hline
			{Exception types and type alias} & Type alias (\ref{subsubsec:type_alias}) & {\color{blue}\ding{108}} & Kotlin's exception types are defined as type aliases to Java's exception types. \\
			
			\hline
			{Top-level functions and top-level properties} & Top-level members (\ref{subsubsec:top-level-members}) & {\color{blue}\ding{108}} &  \\
			
			\hline
			{Package} & Package (\ref{subsubsec:package-open-prob}) & {\color{purple}\ding{74}} & This challenge can also be solved in the component that integrates the analysis with the IDE.  \\
			
			\hline
			{Constructor with default arguments and function with default arguments} & Default argument (\ref{subsubsec:default-arguments}) & {\color{teal}\ding{72}} &  \\
			
			\hline
			{Internal visibility modifier} & Internal modifier (\ref{subsubsec:internal-modifier-open-prob}) & {\color{purple}\ding{74}} &  \\
			
			\hline
			{Sealed class} & Sealed class (\ref{subsubsec:sealed-class-open-prob}) & {\color{purple}\ding{74}} &  \\
			
			\hline
			{Inline class} & Inline class (\ref{subsubsec:inline-class-open-prob}) & {\color{purple}\ding{74}} &  \\
			
			\hline
			{Function returning anonymous object} & Function returning anonymous object (\ref{subsubsec:fun-ret-anonymous-obj-open-prob}) & {\color{purple}\ding{74}} &  \\
			
			\hline
			{Companion object} & Companion object (\ref{subsubsec:companion-object-open-prob}) & {\color{purple}\ding{74}} &  \\
			
			\hline
			{Infix function} & Infix function (\ref{subsubsec:infix-function}) & {\color{blue}\ding{108}} &  \\
			
			\hline
			{Local functions} & Local functions (\ref{subsubsec:local-functions-open-prob}) & {\color{purple}\ding{74}} &  \\
			
			\hline
			{Qualified \keyword{this} object} & qualified \keyword{this} object in  extensions as members (\ref{subsubsec:extensions}), qualified \keyword{this} object in function with receiver type (\ref{subsubsec:higher-order-function-open-prob}) & {\color{purple}\ding{74}} & qualified \keyword{this} object in nested / inner class is same as Java's qualified \keyword{this} in nested / inner class. Therefore, there is no challenge in this scenario \\
			
			\hline
			{Destructuring declaration} & Destructuring declaration (\ref{subsubsec:destructuring-dec-open-prob}) & {\color{purple}\ding{74}} &  \\
			
			\hline
			{Properties accessors} & Properties accessors (\ref{subsubsec:property}) & {\color{blue}\ding{108}} &  \\
			
			\hline
			{Extension function, extension property, companion object extension  and extensions as members} & Extensions (\ref{subsubsec:extensions}) & {\color{purple}\ding{74}} &  \\
			
			\hline
			{Data class} & Destructuring declaration (\ref{subsubsec:destructuring-dec-open-prob}),
			default argument (\ref{subsubsec:default-arguments}) & {\color{purple}\ding{74}}, {\color{teal}\ding{72}} & For a data class, the Kotlin compiler automatically generates the \funRef{componentN} function for the destructuring declaration {\color{purple}\ding{74}}. Additionally, it also generates the \funRef{copy} function with the default value. Therefore, this feature also has the challenge of default argument for the \funRef{copy} function {\color{teal}\ding{72}}. \\
			
			\hline
			{lambda expression, anonymous function, function literal with a receiver, callable reference, class implementing function types} & Higher-order function (\ref{subsubsec:higher-order-function-open-prob}), function type (\ref{subsubsec:data_type}) & {\color{purple}\ding{74}} &  \\
			
			\hline
			{inline function} & Inline function (\ref{subsubsec:inline-fun-open-prob}) & {\color{purple}\ding{74}} &  \\
			
			\hline
			{Operator overloading} & Operator overloading (\ref{subsubsec:operator-overloading}) & {\color{blue}\ding{108}} &  \\
			
			\hline
			{Ranges and Progressions} & Top-level members (\ref{subsubsec:top-level-members}), infix function (\ref{subsubsec:infix-function}), extensions (\ref{subsubsec:extensions}) & {\color{blue}\ding{108}} & In Ranges and Progressions, the methods \funRef{until}, \funRef{downto}, and \funRef{step} are defined as top-level, extension, infix function. \\
			
			\hline
			{Collections and Iterators} & Data type mapping (\ref{subsubsec:data_type}), Top-level members (\ref{subsubsec:top-level-members}), extensions (\ref{subsubsec:extensions}) and destructuring declarations (\ref{subsubsec:destructuring-dec-open-prob}) & {\color{blue}\ding{108}} and {\color{purple}\ding{74}} & Kotlin uses other features such as extensions, destructuring declaration, etc., to define the members of collections and iterators. In addition, some of the collection types are mapped to Java's collection types, and some collection types are defined as type aliases to Java's collection types.  \\
			
			\hline
			
			\multicolumn{4}{c}{} \\
			\multicolumn{4}{c}{Legend} \\
			\multicolumn{4}{c}{{\color{blue}\ding{108}}: Engineering challenge(s) can be solved in the DSL component.}\\
			%			\multicolumn{3}{c}{{\color{atomictangerine}\ding{110}}: Problem(s) can be solved in the  component.} \\
			\multicolumn{4}{c}{{\color{teal}\ding{72}}: Engineering challenge(s) can be solved in the analysis component.} \\
			\multicolumn{4}{c}{{\color{purple}\ding{74}}: Engineering challenge(s) can be solved either in the DSL or analysis component (depends on the analysis designer decision).} \\
		\end{tabular}
		
	\end{adjustbox}
	\caption{List of Kotlin's features that requires an extension in the DSL or analysis components of the existing Java taint analysis tools to support taint analysis for Kotlin programs.}
	\label{table:evaluation-problem-table}
	
\end{table*}

Table \ref{table:evaluation-no-problem-table} summarizes the features of Kotlin that can be analyzed by the existing Java taint analysis tools. {\color{darkgreen}\ding{52}} represents Kotlin's features, for which the Kotlin compiler generates the Java bytecode similar to the respective features in Java. %Hence, the existing Java taint analysis tools have no problem in analyzing these features. 
* represents Kotlin's features, for which the Kotlin compiler generates the Java bytecode similar to some features (third column in the table) in Java, which are not visible in Kotlin source code. For example, the explicit conversion from the \keyword{Number} type to the \keyword{Int} in Kotlin is performed using the method \funRef{toInt}. However, in some scenarios, the Kotlin compiler uses the \funRef{intValue} method in the Java bytecode. Furthermore, the Kotlin compiler uses the \keyword{StringBuilder} \funRef{append} method for the String template, and in some scenarios, it uses Kotlin's \funRef{stringPlus} method. Therefore, we recommend using Java's general propagator methods while analyzing Kotlin programs. {\color{atomictangerine}\ding{52}} represents Kotlin's features, for which the Kotlin compiler generates the Java bytecode similar to the respective features in Java. However, the naming scheme of the generated wrapper class by the Kotlin compiler is different compared to that of the Java compiler. For the default implementation in an interface, the Kotlin compiler generates the Java bytecode differently from Java {\color{red}\ding{52}}. The Java compiler keeps the default implementation in the interface. However, the Kotlin compiler keeps only the abstract methods in the interface, and the default implementation is placed in the generated wrapper class. For example, suppose a developer uses the default implementation method in a class, which implements that interface. In that case, the Kotlin compiler overrides that method automatically and calls the default implementation present in the wrapper class. Whenever that default method is called, the Kotlin compiler calls the virtual method from the object of the derived class or interface similar to the Java compiler. Therefore, there is no engineering challenge with this feature in the DSL, analysis, and IR generator component. However, there may be some challenges with this feature in other components such as the call graph generator component. \\[0.1em]
\customlabel{rqs:rq2}{}
\noindent
\RQ{2:} \textit{For which Kotlin's features, the existing Java taint analysis tools need an extension to support taint analysis for Kotlin programs?}

To answer this research question, we list the Kotlin's features for which the Kotlin compiler generates the Java bytecode differently from the Java compiler. Such differences makes an engineering challenges in the existing Java taint analysis tools analyzing Kotlin programs, as discussed in Section \ref{sec:findings}. Static analysis developers must handle these challegens in the DSL, analysis, or IR generator components.

Table \ref{table:evaluation-problem-table} summarizes the features of Kotlin that require an extension in the existing Java taint analysis tools to support taint analysis for Kotlin. The engineering challenges associated with Kotlin's features are given in the second column. If a Kotlin's feature can be handled in the DSL component, then we categorize that feature into {\color{blue}\ding{108}}. Furthermore, suppose a challenge can be solved in the analysis component without any input from the users of taint analysis tools. In that case, we categorize that feature into {\color{teal}\ding{72}}. For instance, we can solve the default argument challenge (Sub-Section \ref{subsubsec:default-arguments}) in the analysis component without implementing additional constructs in the DSL component.

We did not propose a solution for the challenges discussed in Section \ref{subsec:found_prob_as_open_issues}. However, we can handle these challenges in the DSL or analysis component based on the solution and the analysis designer's decision. Kotlin's features associated with these challenges are categorized into {\color{purple}\ding{74}}. Additionally, for all the features of Kotlin that we manually examined in this exploratory research, Soot generates the valid Jimple code.

\section{\kotlinsecucheck}
\label{sec:secucheck-kotlin-tool}

As a proof of concept, we extended an existing Java taint analysis tool called \secucheck \cite{secucheck} by implementing the solution for six of the engineering challenges discussed in Sub-Section \ref{subsec:found_prob_with_proposed_sol}. For the taint analysis, \secucheck uses Jimple IR \cite{jimple} generated by \soot \cite{soot_paper}. Furthermore, \secucheck provides a DSL called \fluenttql \cite{fluenttql} for specifying taint flows. First, we discuss the implementation of \kotlinsecucheck in Sub-Section \ref{subsec:secu-kot-implementaion}. Then, we evaluate the applicability of \kotlinsecucheck in Sub-Section \ref{subsec:secu-kot-evaluation}.

\subsection{Implementation}
\label{subsec:secu-kot-implementaion}

Table \ref{table:extension-table} summarizes the list of challenges that we handled in \kotlinsecucheck. We implemented the solutions for these challenges without modifying the existing architecture of \secucheck.

\begin{table}[!htbp]
	\scalebox{0.60}{
		\normalsize
		\renewcommand{\arraystretch}{1.1}% Wider
		\begin{tabular}{ m{.15\textwidth} | >{\centering}m{.08\textwidth} | >{\centering\arraybackslash}m{0.50\textwidth}}
			
			\hline
			\multicolumn{1}{c|}{\textbf{Challenges}} & \textbf{Solved in} & \textbf{Newly added constructs in \fluenttql} \\
			
			\hline
			{Data type mapping \newline (\ref{subsubsec:data_type})} & {\color{blue}\ding{108}}* & - \\
			
			\hline
			{Type alias \newline (\ref{subsubsec:type_alias})} & {\color{blue}\ding{108}} & \keyword{TypeAliases} class for experts in Kotlin and domain-experts in custom libraries. The object of \keyword{TypeAliases} are accepted in \keyword{MethodSignatureBuilder}, \keyword{MethodSelector}, and \keyword{MethodConfigurator}. \\ 
			
			\hline
			{Property \newline (\ref{subsubsec:property})} & {\color{blue}\ding{108}} & \funRef{property}, \funRef{getter}, and \funRef{setter} methods \\
			
			\cline{1-3}
			{Top-level members \newline (\ref{subsubsec:top-level-members})} & {\color{blue}\ding{108}} & \funRef{topLevelMember} method \\
			
			\cline{1-3}
			{Extensions \newline (\ref{subsubsec:extensions})} & {\color{blue}\ding{108}} & \funRef{extensionFunction} and \funRef{extensionProperty} methods. For handling qualified \keyword{this} object challenge, provides constants \keyword{QualifiedThis.DISPATCH\_RECEIVER} and \keyword{QualifiedThis.EXTENSION\_RECEIVER} \\
			
			\hline
			{Default argument (\ref{subsubsec:default-arguments})} & {\color{teal}\ding{72}} & - \\
			
			\hline
			
			\multicolumn{3}{c}{} \\ 
			\multicolumn{3}{c}{Legend} \\ 
			\multicolumn{3}{c}{{\color{blue}\ding{108}}: solved in \fluenttql DSL.}\\
			\multicolumn{3}{c}{{\color{blue}\ding{108}}*: solved in \fluenttql without implementing new construct in \fluenttql DSL.}\\
			%			\multicolumn{3}{c}{{\color{atomictangerine}\ding{110}}: Problem(s) can be solved in the  component.} \\
			\multicolumn{3}{c}{{\color{teal}\ding{72}}: solved in the analysis component.} \\
		\end{tabular}
		
	}
	\caption{List of found engineering challenges handled in \kotlinsecucheck.}
	\label{table:extension-table}
	\vspace{-4mm}
\end{table}

For handling the data type mapping discussed in Sub-Section \ref{subsubsec:data_type}, we implemented a data type transformer module in \fluenttql \hspace{0.001mm} {\color{blue}\ding{108}}*. This transformer checks whether the given type in a method signature is a valid Kotlin type or function type, as described in Tables \ref{tab:functiontypesmapping} and \ref{tab:datatypesmapping}, and transforms the given type into respective Java data type. This allows users to provide Kotlin types such as \keyword{kotlin.Int}, \keyword{kotlin.Int?}, etc., in a method signature. In addition, users can also provide short type names such as \keyword{Int}, \keyword{Int?}, etc. Furthermore, for a function type, users can provide a regular expression such as ``\constant{() $\rightarrow$ \_}'', or a function type itself such as ``\constant{() $\rightarrow$ String}''. The transformer looks for the function type expression and transforms it into a valid data type, as summarized in Table \ref{tab:functiontypesmapping}. The limitation of this current implementation is that users can not provide complex function types such as \constant{(Int) $\rightarrow$ (Int, Int) $\rightarrow$ String}. For such function types, users must use regular expressions, e.g., \constant{(\_) $\rightarrow$ \_}. Additionally, suppose users want to specify to track a parameter of function type in \fluenttql. In that case, users must explicitly specify the propagation rules for the \funRef{invoke} method of the \keyword{Function*} class as discussed in Sub-Section \ref{subsubsec:higher-order-function-open-prob}.

For handling type alias (\ref{subsubsec:type_alias}), property (\ref{subsubsec:property}), top-level members (\ref{subsubsec:top-level-members}), and extensions (\ref{subsubsec:extensions}), we implemented new constructs in \fluenttql that helps the users to specify the respective features \hspace{0.001mm} {\color{blue}\ding{108}}. Listing \ref{lst:secu-kotlin-example} demonstrates the way of specifying type aliases and extension property in \fluenttql of \kotlinsecucheck. For the type alias challenge, we implemented the \keyword{TypeAliases} class in \fluenttql, which experts in Kotlin programming language or the domain experts in custom libraries can use to specify type aliases. For instance, experts in the Kotlin programming language can specify the type aliases defined in the Kotlin standard library as shown in Lines 2-6. Then, the users of \fluenttql can use the specified type aliases in \keyword{MethodConfigurator} (Line 14), \keyword{MethodSelector}, or \keyword{MethodSignatureBuilder}, which replaces the given type alias with the original type as specified by the experts. Note: \secucheck provides \keyword{MethodSignatureBuilder} for novice users to build a method signature with fluent interface. Similarly, it provides \keyword{MethodConfigurator} and \keyword{MethodSelector} for configuring methods with taint information using fluent interface.

For handling the property, top-level members, and extensions, we provide the methods in the fluent interface of \keyword{MethodSignatureBuilder}. For example, for properties, the methods are \funRef{property}, \funRef{getter} (Line 12), and \funRef{setter}. If a property is an extension, then the method is \funRef{extensionProperty} (Line 11). This function takes three arguments---receiver type, property name, and property type. From these inputs, \fluenttql builds the valid method signature. Similarly, \funRef{extensionFunction} method for specifying extension functions and the \funRef{topLevelMember} method for specifying top-level members. For handling the qualified \keyword{this}-object in extensions, we provide two constants---\keyword{Qualified.DISPATCH\_RECEIVER} and \keyword{Qualified.EXTENSION\_RECEIVER}, which can be used in the method \funRef{thisObject} (Line 15) of \keyword{MethodConfigurator} to track the respective \keyword{this} object.
The limitation for this implementation is that these methods are only available in the method chain of \keyword{MethodSignatureBuilder} and not available for \keyword{MethodConfigurator} and \keyword{MethodSelector}. Similarly, the qualified \keyword{this} constants are only available for \keyword{MethodConfigurator}, and it is not available for \keyword{MethodSelector}. Finally, we handled the challenge of default argument in the analysis component {\color{teal}\ding{72}}, as proposed in Sub-Section \ref{subsubsec:default-arguments}.

\begin{lstlisting}[float=*,caption={Example of type alias, extension property in \fluenttql of \kotlinsecucheck.}, style=Java,captionpos=b,label={lst:secu-kotlin-example}, abovecaptionskip=1pt, belowcaptionskip=-13pt, basicstyle=\tt\fontfamily{pcr}\selectfont\footnotesize, escapechar=!]
	// Specified by Kotlin programming language experts.
	static TypeAliases typeAliases = new TypeAliases(){{
					add("ArrayList", "java.util.ArrayList");
					add("HashSet", "java.util.HashSet");
					...
	}};
	
	// Specified by the users of fluentTQL
	public MethodSignature signature = new MethodSignatureBuilder()
									.atClass("de.fraunhofer.iem.EmployeePrinter")
									.extensionProperty("de.fraunhofer.iem.Employee", "nameLength", "Int")
									.getter();
	
	public Method source1 = new MethodConfigurator(signature, typeAliases)
									.in().thisObject(QualifiedThis.DISPATCH_RECEIVER)
									.out().returnValue()
									.configure();
\end{lstlisting}
\subsection{Evaluation}
\label{subsec:secu-kot-evaluation}

To evaluate the applicability of \kotlinsecucheck, we found a vulnerable version of the Spring PetClinic application written in Kotlin\footnote{\url{https://shorturl.at/hvyRS}}. This project contains 27 Kotlin files with six known hibernate injections as summarized in Table \ref{table:analysisResultOverview}. 

\begin{table}[!h]
	\begin{adjustbox}{max width=\columnwidth}
		\renewcommand{\arraystretch}{1}% Wider
		\normalsize
		\begin{tabular}{>{\centering}m{0.23\textwidth}|c|c|c|c|c|c|c|c}
			\hline
			
			\textbf{Project Name} & \rotatebox{75}{\parbox{2.55cm}{{\texttt{\#}Kotlin-files}}} & \rotatebox{75}{{\texttt{\#}Taint-flows}} & 
			\rotatebox{75}{{\texttt{\#}Queries}} & 
			\rotatebox{75}{\texttt{\#}Found-flows} & 
			\rotatebox{75}{\texttt{\#}Runtime (s)} &
			\rotatebox{75}{\parbox{3cm}{Display\\ error messages?}} &  
			\rotatebox{75}{\parbox{3cm}{Display\\ line numbers?}} & 
			\rotatebox{75}{\parbox{3cm}{Display\\ file locations?}}\\
			
			\hline
			
			spring-petclinic-kotlin\newline(vulnerable) & 
			27 & 
			6 & 
			5 & 
			6 & 
			11.05 & 
			{\color{darkgreen}\ding{52}} & 
			{\color{darkgreen}\ding{52}} & 
			{\color{darkgreen}\ding{52}} \\ %\hline
			
			\hline
			
			\multicolumn{9}{c}{} \\
			\multicolumn{9}{c}{\texttt{\#}Kotlin-files: Number of Kotlin files in the project} \\
			\multicolumn{9}{c}{\texttt{\#}Taint-flows: Number of known taint-flows in the project} \\ 
			\multicolumn{9}{c}{\texttt{\#}Queries: Number of specified taint-flow queries in \fluenttql of \kotlinsecucheck} \\
			\multicolumn{9}{c}{\texttt{\#}Found-flows: Number of found taint-flows by \kotlinsecucheck} \\
			\multicolumn{9}{c}{\texttt{\#}Runtime (s): Runtime in seconds (average of 10 runs)} \\
			
		\end{tabular}
		
	\end{adjustbox}
	\caption{Overview of \kotlinsecucheck analysis results.}
	\label{table:analysisResultOverview}
\end{table}

\kotlinsecucheck found all the six taint-flows with the run time of 11.05 seconds (average of 10 runs). \kotlinsecucheck successfully displayed the valid line numbers of the source and sink methods. It also displayed the customized error message as well as the descriptive messages from the \fluenttqltoEng translator \cite{secucheck}. Additionally, \kotlinsecucheck displayed the file locations of the source and sink methods. However, \secucheck through the command prompt displays the file location of the classes instead of the Java files in the Static Analysis Results Interchange Format (SARIF) \cite{sarif} output. Therefore, \kotlinsecucheck has no problem in displaying the valid file locations. However, suppose developers want to display the file location of the Kotlin files instead of the class files in the SARIF output. In that case, file location of the Kotlin files has to be identified based on the challenge we discussed in Section \ref{sec:findings}.

\section{Conclusion and Future work}
\label{sec:conclusion}
In this paper, we presented our exploratory study for Kotlin taint analysis, which shows that most of the Kotlin constructs can be analyzed by an existing Java taint analysis tool. However, we found 18 engineering challenges that must be handled differently than the Java taint analysis. For eight of these challenges, we proposed solutions. Finally, as a proof of concept, we extended an existing Java taint analysis, \secucheck, by implementing six of these solutions, which led to \kotlinsecucheck. We evaluated the applicability of \kotlinsecucheck, which found all the six expected taint-flows. In the future, we plan to work on the open issues from Sub-Section~\ref{subsec:found_prob_as_open_issues} and extend the implementation of \kotlinsecucheck, after which a thorough evaluation with real-world applications can be performed.

\bibliographystyle{IEEEtran}
\bibliography{bibliography}

\begin{thebibliography}{10}
\providecommand{\url}[1]{#1}
\csname url@samestyle\endcsname
\providecommand{\newblock}{\relax}
\providecommand{\bibinfo}[2]{#2}
\providecommand{\BIBentrySTDinterwordspacing}{\spaceskip=0pt\relax}
\providecommand{\BIBentryALTinterwordstretchfactor}{4}
\providecommand{\BIBentryALTinterwordspacing}{\spaceskip=\fontdimen2\font plus
\BIBentryALTinterwordstretchfactor\fontdimen3\font minus
  \fontdimen4\font\relax}
\providecommand{\BIBforeignlanguage}[2]{{%
\expandafter\ifx\csname l@#1\endcsname\relax
\typeout{** WARNING: IEEEtran.bst: No hyphenation pattern has been}%
\typeout{** loaded for the language `#1'. Using the pattern for}%
\typeout{** the default language instead.}%
\else
\language=\csname l@#1\endcsname
\fi
#2}}
\providecommand{\BIBdecl}{\relax}
\BIBdecl

\bibitem{cwe89}
``{CWE-89: Improper Neutralization of Special Elements used in an SQL
  Command},'' \url{https://cwe.mitre.org/data/definitions/89.html}, accessed:
  2021-June-22.

\bibitem{ktlint}
``{\textsc{ktlint: }An anti-bikeshedding Kotlin linter with built-in
  formatter},'' \url{https://github.com/pinterest/ktlint}, accessed:
  2021-December-14.

\bibitem{detekt}
``{\textsc{detekt: }static analysis for Kotlin},''
  \url{https://github.com/detekt/detekt}, accessed: 2021-December-14.

\bibitem{diktat}
``{\textsc{diktat: }Strict coding standard for Kotlin and a custom set of rules
  for detecting code smells, code style issues and bugs},''
  \url{https://github.com/diktat-static-analysis/diKTat}, accessed:
  2021-December-14.

\bibitem{sonar}
``{\textsc{SonarQube: }automatic code review tool to detect bugs,
  vulnerabilities, and code smells},''
  \url{https://docs.sonarqube.org/latest/}, accessed: 2021-December-14.

\bibitem{sonarRules}
``{\textsc{SonarQube: } rules for Kotlin.}''
  \url{https://rules.sonarsource.com/kotlin}, accessed: 2021-December-14.

\bibitem{fluenttql}
\BIBentryALTinterwordspacing
G.~Piskachev, J.~Sp{\"{a}}th, I.~Budde, and E.~Bodden, ``Fluently specifying
  taint-flow queries with fluenttql,'' \emph{Empir. Softw. Eng.}, vol.~27,
  no.~5, p. 104, 2022. [Online]. Available:
  \url{https://doi.org/10.1007/s10664-022-10165-y}
\BIBentrySTDinterwordspacing

\bibitem{cwe77}
``{CWE-77: Improper Neutralization of Special Elements used in a Command
  ('Command Injection')},''
  \url{https://cwe.mitre.org/data/definitions/77.html}, accessed:
  2021-October-25.

\bibitem{cwe476}
``{CWE-476: NULL Pointer Dereference},''
  \url{https://cwe.mitre.org/data/definitions/476.html}, accessed:
  2021-June-18.

\bibitem{xss}
D.~Endler, ``The evolution of cross site scripting attacks,'' Technical report,
  iDEFENSE Labs, Tech. Rep., 2002.

\bibitem{flowdroid}
S.~Arzt, S.~Rasthofer, C.~Fritz, E.~Bodden, A.~Bartel, J.~Klein, Y.~Le~Traon,
  D.~Octeau, and P.~McDaniel, ``Flowdroid: Precise context, flow, field,
  object-sensitive and lifecycle-aware taint analysis for android apps,''
  \emph{Acm Sigplan Notices}, vol.~49, no.~6, pp. 259--269, 2014.

\bibitem{secucheck}
G.~Piskachev, R.~Krishnamurthy, and E.~Bodden, ``Secucheck: Engineering
  configurable taint analysis for software developers,'' in \emph{2021 IEEE
  21st International Working Conference on Source Code Analysis and
  Manipulation (SCAM)}, 2021, pp. 24--29.

\bibitem{soot}
P.~Lam, E.~Bodden, O.~Lhot{\'a}k, and L.~Hendren, ``{The Soot framework for
  Java program analysis: a retrospective},'' in \emph{Cetus Users and Compiler
  Infastructure Workshop (CETUS 2011)}, vol.~15, no.~35, 2011.

\bibitem{kotlinOfficialDoc}
``{Kotlin's official documentation},''
  \url{https://kotlinlang.org/docs/home.html}, accessed: 2021-November-21.

\bibitem{jetbrainsStatistics}
``{The State of Developer Ecosystem 2020},''
  \url{https://www.jetbrains.com/lp/devecosystem-2020/kotlin/}, accessed:
  2021-October-27.

\bibitem{jimple}
R.~Vallee-Rai and L.~J. Hendren, ``{Jimple: Simplifying Java bytecode for
  analyses and transformations},'' 1998.

\bibitem{soot_paper}
\BIBentryALTinterwordspacing
R.~Vall{\'e}e-Rai, P.~Co, E.~Gagnon, L.~Hendren, P.~Lam, and V.~Sundaresan,
  ``Soot: A java bytecode optimization framework,'' in \emph{CASCON First
  Decade High Impact Papers}, ser. CASCON '10.\hskip 1em plus 0.5em minus
  0.4em\relax Riverton, NJ, USA: IBM Corp., 2010, pp. 214--224. [Online].
  Available: \url{https://doi.org/10.1145/1925805.1925818}
\BIBentrySTDinterwordspacing

\bibitem{sarif}
S.~Kummita and G.~Piskachev, ``Integration of the static analysis results
  interchange format in cognicrypt,'' \emph{arXiv preprint arXiv:1907.02558},
  2019.

\end{thebibliography}

\end{document}